# Mass asymmetry effects on geometry of vanishing flow.

Supriya Goyal and Rajeev K. Puri*
*Department of Physics, Panjab University, Chandigarh-160014, INDIA*
*\* email: rkpuri@pu.ac.in*

## Introduction

In the heavy-ion collisions at intermediate energies, one can see various phenomena like onset of multifragmentation [1], collective flow and its disappearance [2] as well as a mixture of fusion, decay and fission [1]. Among all these phenomena, collective flow enjoys a special place due to its sensitivity towards the nuclear matter equation of state as well as towards nucleon-nucleon cross-section [2]. The collective flow is also found to be highly sensitive toward the size of the system [3], the incident energy of the projectile [4], and the colliding geometry (i.e., the impact parameter) [5]. In a recent study by Puri *et al.* [6], it has been found that while going from perfectly central collisions to most peripheral ones, the collective transverse flow passes through a maximum (at low values of impact parameter), a zero value (at some intermediate value of impact parameter), and achieves negative values (at large values of impact parameter), at a fixed incident energy. The intermediate value of impact parameter where collective transverse flow vanishes is termed as Geometry of Vanishing Flow (GVF). It has been found that mass dependence of GVF is insensitive to the nuclear matter equation of state and momentum dependent interactions, whereas it is quite sensitive to the nucleon-nucleon cross-section. It is noted that the study was done only for the symmetric systems. Therefore, in the present study we aim to find the role of mass asymmetry on the GVF at a fixed value of energy. The present calculations are done with Quantum Molecular Dynamics (QMD) model [7].

## Model

The QMD model simulates the heavy-ion reactions on event by event basis. This is based on a molecular dynamic picture where nucleons interact via two and three-body interactions. The nucleons propagate according to the classical equations of motion:

$$\frac{d\mathbf{r}_i}{dt} = \frac{dH}{d\mathbf{p}_i} \text{ and } \frac{d\mathbf{p}_i}{dt} = -\frac{dH}{d\mathbf{r}_i}, \quad (1)$$

where H stands for the Hamiltonian which is given by

$$H = \sum_i \frac{\mathbf{p}_i^2}{2m_i} + V^{tot}. \quad (2)$$

Our total interaction potential $V^{tot}$ reads as

$$V^{tot} = V^{Loc} + V^{Yuk} + V^{Coul} + V^{MDI}, \quad (3)$$

where $V^{Loc}$, $V^{Yuk}$, $V^{Coul}$, and $V^{MDI}$ are, respectively, the local (two and three-body) Skyrme, Yukawa, Coulomb and momentum dependent potentials. $E_{bal}$ is calculated by using the quantity "*directed transverse momentum* $<P^{dir}_x>$", which is defined as:

$$\left\langle p_x^{dir} \right\rangle = \frac{1}{A}\sum_{i=1}^{A} sign\{y(i)\} p_x(i). \quad (4)$$

Here $y(i)$ is the rapidity and $p_x(i)$ is the transverse momentum of $i^{th}$ particle. The rapidity is defined as:

$$y(i) = \frac{1}{2}\ln\frac{E(i) + p_z(i)}{E(i) - p_z(i)}, \quad (5)$$

where $E(i)$ and $p_z(i)$ are, respectively, the total energy and longitudinal momentum of $i^{th}$ particle. The $E_{bal}$ was then deduced using a straight line interpolation.

## Results and discussion

The asymmetry of a reaction is defined by the parameter called asymmetry parameter ($\eta$) and is given by:

$$\eta = \left|\frac{A_T - A_P}{A_T + A_P}\right|, \qquad (6)$$

where $A_T$ and $A_P$ are the masses of target and projectile, respectively. The $\eta = 0$ corresponds to symmetric reactions and nonzero values of $\eta$ defines different asymmetries of a reaction. For the present study, we simulated the various reactions for 1000-5000 events, at full range of colliding geometries ranging from the central to peripheral collisions in small steps of 0.25 and at fixed incident energy of 200 MeV/nucleon. The asymmetry of a reaction is varied from $\eta = 0$-0.9, keeping the total mass of the system ($A_{TOT}$) fixed as 40, 80, 160 and 240. For the present study, we employed a soft equation of state (K=200 MeV) with momentum dependent interactions along with energy dependent cugnon cross-section. In fig. 1, we display GVF as a function of $\eta$ for $A_{TOT} = 40$, 80, 160, and 240. Symbols are explained in the caption of the figure. Lines are just to guide the eye. The percentage variation in GVF while going from $\eta = 0$ to 0.9 is −62%, −42.03%, −40.91%, and −21.21%, respectively, for $A_{TOT} = 40$, 80, 160, and 240. The negative signs indicate that GVF decreases with increase in $\eta$. It is very clear from the figure that with increase in system mass, the effect of mass asymmetry of the reaction on GVF decreases. It is well known that the energy at which flow vanishes i.e. $E_{bal}$, increases with increase in $\eta$ and impact parameter, while with increase in $A_{TOT}$, it decreases. This is due to the decrease in nn collisions with increase in $\eta$ and impact parameter, and increase in Coulomb repulsion with increase in $A_{TOT}$. The present study has been done at fixed incident lab energy; therefore, the effective centre-of-mass energy also decreases as $\eta$ increases for every fixed system mass. To compensate all these factors, the value of impact parameter, where flow vanishes, decreases as $\eta$ increases.

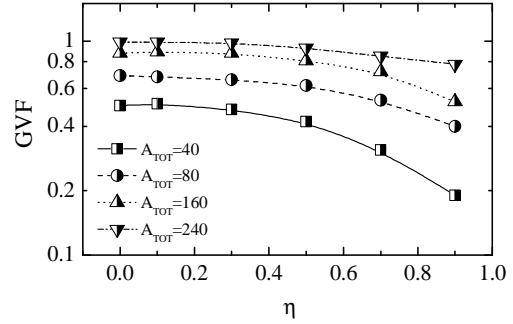

**Fig. 1** The geometry of vanishing flow (GVF) as a function of $\eta$ for different system masses. The results for different system masses ($A_{TOT}$) = 40, 80, 160, and 240 are represented, respectively, by the half filled squares, circles, triangles, and inverted triangles..

## Acknowledgments

This work is supported by a research grant from the Council of Scientific and Industrial Research (CSIR), Govt. of India, vide grant No. 09/135(0563)/2009-EMR-1.